\documentclass[conference]{IEEEtran}
\IEEEoverridecommandlockouts

\usepackage{cite}
\usepackage{multirow}
\usepackage{url}
\usepackage{graphicx}
\usepackage{amsmath}
\usepackage{amsfonts}
\usepackage{caption}
\usepackage{subcaption}

 \newcommand{\squishlist}{
	\begin{list}{$\bullet$}
		{ \setlength{\itemsep}{0pt}
			\setlength{\parsep}{3pt}
			\setlength{\topsep}{3pt}
			\setlength{\partopsep}{0pt}
			\setlength{\leftmargin}{1.5em}
			\setlength{\labelwidth}{1em}
			\setlength{\labelsep}{0.5em} } }
	\newcommand{\squishlisttwo}{
		\begin{list}{$\bullet$}
			{ \setlength{\itemsep}{0pt}
				\setlength{\parsep}{0pt}
				\setlength{\topsep}{0pt}
				\setlength{\partopsep}{0pt}
				\setlength{\leftmargin}{2em}
				\setlength{\labelwidth}{1.5em}
				\setlength{\labelsep}{0.5em} } }
		\newcommand{\squishend}{
	\end{list}  }
 
\def\BibTeX{{\rm B\kern-.05em{\sc i\kern-.025em b}\kern-.08em
    T\kern-.1667em\lower.7ex\hbox{E}\kern-.125emX}}
\begin{document}

\title{ID and Graph View Contrastive Learning with Multi-View Attention Fusion for Sequential Recommendation }

\author{\IEEEauthorblockN{Xiaofan Zhou}
\IEEEauthorblockA{
\textit{Worcester Polytechnic Institute}\\
Worcester, USA \\
xzhou5@wpi.edu}
\and
\IEEEauthorblockN{Kyumin Lee}
\IEEEauthorblockA{
\textit{Worcester Polytechnic Institute}\\
Worcester, USA \\
kmlee@wpi.edu}}



\maketitle
\begin{abstract}
Sequential recommendation has become increasingly prominent in both academia and industry, particularly in e-commerce. The primary goal is to extract user preferences from historical interaction sequences and predict items a user is likely to engage with next. Recent advances have leveraged contrastive learning and graph neural networks to learn more expressive representations from interaction histories --- graphs capture relational structure between nodes, while ID-based representations encode item-specific information. However, few studies have explored multi-view contrastive learning between ID and graph perspectives to jointly improve user and item representations, especially in settings where only interaction data is available without auxiliary information.
To address this gap, we propose \textbf{M}ulti-\textbf{V}iew \textbf{C}ontrastive learning for sequential \textbf{rec}ommendation (\textbf{MVCrec}), a framework that integrates complementary signals from both sequential (ID-based) and graph-based views. MVCrec incorporates three contrastive objectives: within the sequential view, within the graph view, and across views. To effectively fuse the learned representations, we introduce a multi-view attention fusion module that combines global and local attention mechanisms to estimate the likelihood of a target user purchasing a target item. Comprehensive experiments on five real-world benchmark datasets demonstrate that MVCrec consistently outperforms 11 state-of-the-art baselines, achieving improvements of up to 14.44\% in NDCG@10 and 9.22\% in HitRatio@10 over the strongest baseline. Our code and datasets are available at \url{https://github.com/sword-Lz/MMCrec}.
\end{abstract}



\section{Introduction}
Sequential recommendation has received increasing attention from both industry and academia, with the primary focus being on recommending items based on users' chronologically ordered purchase histories \cite{liu2023joint,liang2023mmmlp,Lin_2023,li2023automlp,wang2020fine,Yang_2022, 10.1145/3580305.3599519}. In the early stage, researchers applied recurrent neural network (RNN) and convolutional neural network (CNN) to sequential recommendation \cite{DBLP:journals/corr/HidasiKBT15,tang2018personalized,kang2018self}. Additionally, self-supervised methods have been employed in sequential recommendation; for example, BERT4Rec \cite{sun2019bert4rec} utilizes BERT as an encoder for sequential lists. More recently, contrastive learning and related techniques have been adopted in sequential recommendation to enhance the effectiveness of learned representations \cite{xie2022contrastive, Qiu_2022, Qin_2023}.

However, the utilization of contrastive learning (CL) to effectively capture the information of historical sequences remains a challenging research area. Contrastive learning aims to maximize the dissimilarity between different categories of individuals (e.g., users or items) while minimizing the dissimilarity within the same category. The first obstacle often lies in selecting suitable augmentation operations for generating similar instances. To date, three classes of augmentation operations have been established. The first class generates different views of the same sequence through random operations like `masking', `cropping', or `reordering' items \cite{Chen_2022, liu2021contrastive}. The second class uses variable dropout probabilities at the model level to create different views of the same sequential data \cite{Qiu_2022}. The third class combines `neural mask', `layer drop', and `encoder complement' with data augmentation techniques for constructing positive and negative view pairs \cite{liu2022improving}.

Most of the prior works leverage sequence information to perform contrastive learning on individual sequences. They employ data augmentation or model-level augmentation techniques to augment the historical sequences. Subsequently, the InfoNCE objective function \cite{oord2018representation} is utilized to compute the contrastive loss. This objective function aims to minimize the distance between augmented sequences generated from the same original sequence, while maximizing the distance between augmented sequences generated from different original sequences.

Although these prior methods have achieved some effectiveness in sequential recommendation, they are suboptimal because of the neglect of structural information which can be obtained/learned from graph-based methods. Graph-based recommendation systems provide a more comprehensive representation of users and items by fully exploiting graph structures, thereby making significant contributions to the field of recommendation systems. In basic recommendation approaches, NGCF \cite{10.1145/3397271.3401063} and LightGCN \cite{Wang_2019} integrate graph convolutional networks into the recommendation systems. UltraGCN \cite{10.1145/3459637.3482291} simplifies GCNs for collaborative filtering by omitting feature transformations and nonlinear activations. As contrastive learning has developed, VGCL \cite{10.1145/3539618.3591691} employs variational graph reconstruction to estimate the Gaussian distribution of each node and generates multiple contrastive views through multiple samplings from the estimated distributions. CGCL \cite{he2023candidate} explore a new way to build contrastive pairs by using similar semantic embeddings. In the realm of sequential recommendation, graph contrastive learning also plays a significant role; MAErec \cite{Ye_2023} applies graph contrastive learning to adaptively and dynamically distill global item transitional information in self-supervised augmentation scenarios with scarce labels. However, even though many multi-modal recommendation methods \cite{li2024category, li2026cpgrec+} are proposed, cross-view contrastive learning between graph and sequence information remains a less explored area in sequential recommendation, especially when given only interaction data without any auxiliary information. 

To fill the gap, in this paper, we propose a novel framework based on multi-view contrastive learning, named \textbf{M}ulti\textbf{V}iew \textbf{C}ontrastive learning for sequential \textbf{rec}ommendation (MVCrec). Initially, we use contrastive learning to learn each user's historical sequence representation. To make the most of graph structure given the sequence information, we also build an item-based graph and apply contrastive learning to learn the structural representation from the historical sequence. According to common sense, embedding of item IDs provides more item-specific information, whereas utilizing a graph structure to represent items captures more information about their relationships with other items. To further enhance our understanding of structural and sequential representations, we introduce and implement a cross-view contrastive learning strategy. This strategy is designed to pull out more detailed features, generating extra contrastive pairs, which are compared with data-augmented views during the training. Finally, given the two different sequence representations (i.e., item-based sequence representation and graph-based sequence representation) which are created by the contrastive learning, we run our proposed multi-view attention fusion module to combine structural and sequential features. In the experiment, we found that both sequence and graph structures positively contributed to improving the effectiveness, with the graph structure having a greater impact than the sequence view.

In summary, the major contributions of our proposed MVCrec are as follows:
\squishlist
\item We propose a novel multi-view contrastive learning approach in the sequential recommendation domain. 
The proposed model proficiently extracts relevant information from both positive and negative samples by utilizing sequence and graph views derived from users' historical item lists (i.e., prior interaction data). 

\item A multi-view attention fusion module is proposed to be seamlessly integrated into MVCrec to calculate the recommendation score, utilizing representations from diverse views.

\item Through comprehensive experiments across five public benchmark datasets, we demonstrate that MVCrec outperforms 11 state-of-the-art baselines.
\squishend 

\begin{figure*}
  \centering
  \includegraphics[width=1\linewidth]{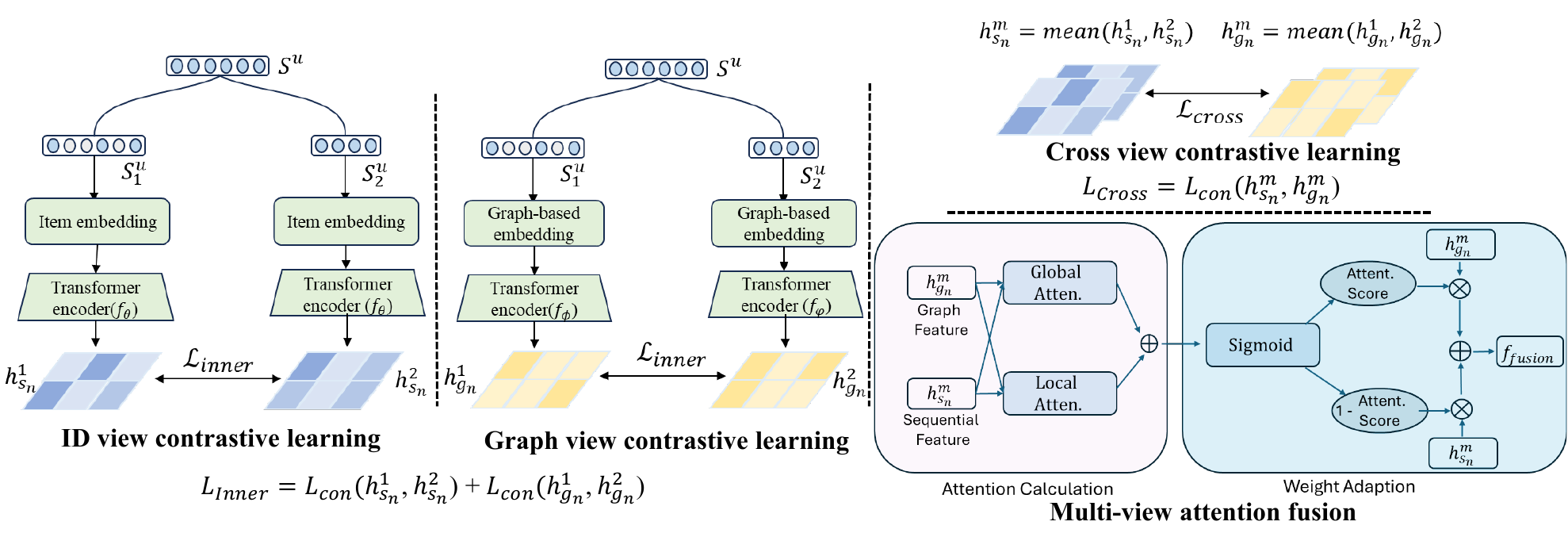}
  \caption{Our proposed framework, MVCrec, consists of multi-view contrastive learning and multi-view attention fusion module.}
  \label{fig:2}
\end{figure*}

\section{Related work}
\subsection{Sequential recommendation} Sequential recommendation is deployed to forecast user preferences based on their historical purchases. In the initial phase of sequential recommendation development, the Markov chain was utilized to formulate predictions by modeling stochastic transitions and uncovering sequential patterns \cite{garcin2013personalized, feng2015personalized}. 

With the growth of deep learning in many areas, RNN and Transformer-based methods have been used in sequential recommendation and have achieved good results. They are good at understanding both the long-term and short-term information in users' historical sequences. For example, GRU4rec \cite{DBLP:journals/corr/HidasiKBT15} uses Gated Recurrent Units (GRU) to learn sequential information from the previously consumed items. Caser \cite{tang2018personalized} uses both horizontal and vertical CNNs to understand sequential behaviors. SASRec \cite{kang2018self} was the first to use the attention mechanism in sequential recommendation. In terms of Transformer, BERT4Rec \cite{sun2019bert4rec} uses deep bidirectional self-attention to understand the possible relationships between items and sequences. LinRec \cite{10.1145/3539618.3591717} introduces a novel method that enhances efficiency while retaining the learning capabilities of traditional dot-product attention through a linear attention module. MELT \cite{Kim_2023} mutually enhance user and item bilateral branches to deal with long-tailed problem. 

While these methodologies have made some advancements in the field of sequential recommendation, most of them have not incorporated structured information, such as graph structures, into their considerations. Unlike the prior works, our approach concurrently uses information derived from both graphs and sequences.

\smallskip
\subsection{Contrastive learning} To enable deep learning models to more accurately differentiate instances pertaining to distinct individuals, contrastive learning was introduced in \cite{Wu_2018_CVPR}. The core concept of contrastive learning is to maximize the dissimilarity between varying individuals, and it has witnessed substantial advancements in recent years. The work of \cite{oord2018representation} introduced the use of mutual information to quantify the similarity between two individuals, considering different views of the same individual as positive pairs. Subsequently, \cite{he2020momentum} employed a queue to manage the extensive dictionary associated with contrastive learning, while \cite{chen2020simple} leveraged the remaining pairs in the batch as the negative pairs for the positive pair, introducing a projector to enhance the performance of contrastive learning further. Additionally, \cite{grill2020bootstrap} explored the execution of contrastive learning tasks without the incorporation of negative samples. In multi-view contrastive learning, MSM4SR \cite{zhang2023multimodal} proposes fusing text and image views before applying contrastive learning. However, this approach overlooks the interrelationship of cross-view contrastive learning. On the other hand, MMSSL \cite{wei2023multi} suggests using GCN for cross-view contrastive learning, but it doesn't account for sequential data.

Recently, contrastive learning has been used in sequential recommendation to handle issues like not having enough data and having data that’s noisy. CL4rec \cite{xie2022contrastive} learns about users by comparing different views of the same sequence data. It uses random actions like ‘mask’, ‘crop’, or ‘reorder’ items to create these different views. DuoRec \cite{Qiu_2022} makes pairs to compare by using ``dropout'' at the model level and suggests using sequences with the same next interaction as matching pairs instead of comparing different data views. MCLrec \cite{Qin_2023} offers a meta-learning strategy to train contrastive learning with the goal to address the problem of sparse data and create more meaningful representations. EMKD \cite{du2023ensemble} proposes knowledge distillation, which uses contrastive learning to facilitate knowledge transfer between parallel networks, and uses the ensemble of different models as the final prediction. DCrec \cite{yang2023debiased} introduces a new global learning strategy to deal with popularity bias in sequential recommendation. Wu et al. \cite{wu2022multi} proposed a Multi-behavior Multi-view Contrastive Learning Recommendation (MMCLR), which uses different behaviors of users as positive pair for recommendation, which requires multi-behavior data for implementation. MCLSR\cite{wang2022multi} proposed multi-level contrastive learning framework for sequential recommendation. Unlike our work utilizing only target behavior with item-based sequence embeddings and graph-based item embeddings, 
MMCLR employs cross-view contrastive learning by constructing various types of graphs to generate representations for users or items. It also neglects the inner structure in different view, which deserves to apply contrastive learning.

In this paper, the principle of contrastive learning is adapted to extract superior representations of historical interaction sequences, and a new multi-view contrastive learning approach which includes inner view and cross view contrastive leaning is proposed. 
\smallskip
\subsection{Graph-based recommendation} User and item interactions in the recommendation task naturally form a graph structure; thus, the incorporation of graph structures is prevalent in recommendation systems. Foundational recommendations like NGCF \cite{10.1145/3397271.3401063} and LightGCN \cite{Wang_2019} have advanced the field of recommendation by integrating GCN structures, thus, enhancing the developmental trajectory of recommendation systems. UltraGCN \cite{10.1145/3459637.3482291} further refines the approach by streamlining GCNs for collaborative filtering and omitting unnecessary feature transformations and nonlinear activations. Additionally, works like CGCL \cite{he2023candidate} and VGCL \cite{10.1145/3539618.3591691} have applied graph structures to contrastive learning, utilizing auto-encoders to optimize the process. SRGNN  \cite{Wu_2019} was proposed to use GNN structure to train the sequential recommendation. Within the realm of sequential recommendation, MAErec \cite{Ye_2023} ingeniously employs graph data in contrastive learning to address issues related to label scarcity. In this paper, we also construct a graph for items to learn their embeddings and user preference representation from the historical sequence via multi-view contrastive learning.

\section{Problem definition}
The primary objective of this paper is to predict the next item, \( c_{n+1} \), which a user \textit{u} is likely to purchase based on the user's historical sequence, denoted as \( S^u = [c_1, c_2, \ldots, c_n] \). In this notation, \( c_i \) represents the \( i \)-th item that the user has purchased, and \( n \) is the length of the user's purchasing history.

\section{Proposed Method}

\subsection{Overview}
As depicted in Figure~\ref{fig:2}, our proposed MVCrec learns two types of item embedding (typical item embedding and graph-based item embedding), and integrates two contrastive learning approaches: graph-based and sequence-based contrastive learning. Each approach consists of a stochastic data augmentation module, a sequence encoder, and a contrastive loss function~\cite{xie2022contrastive}. To optimally leverage information from both graph and sequence data, MVCrec employs a cross-view contrastive loss, complementing the two contrastive learning approaches. Additionally, a multi-view attention fusion module is formulated to amalgamate item-based sequence representation and graph-based sequence representation from both views. In essence, MVCrec consists of five components: (1) stochastic data augmentation module, (2) item embeddings, (3) Transformer-based sequence encoder, (4) multi-view contrastive learning, and (5) multi-view attention fusion module. Detailed information about these modules are described in the following subsections. 

\subsection{Stochastic data augmentation module}
\label{sec:aug}
This module aims to generate two positive views for each historical sequence. Inspired by CL4rec \cite{xie2022contrastive}, we apply three stochastic data augmentations — `masking’, `cropping’, and `reordering’ — to the historical sequence. The procedure for generating two augmented sequences is as follows:

\begin{equation}
\begin{aligned}
\tilde{S}_{1}^{u}=g_{1}\left(S^{u}\right), \tilde{S}_{2}^{u}=g_{2}\left(S^{u}\right) 
\end{aligned}
\end{equation}
where $g_1$ and $g_2$ are a pair of different stochastic data augmentation methods (i.e., randomly select two of `mask', `crop' and `reorder'), and $\tilde{S}_{1}^{u}$ and $\tilde{S}_{2}^{u}$ are a pair of positive samples.

\subsection{Two types of item embedding}
\label{sec:embeddings}
Initially, we project all items into a common embedding space \cite{kang2018self}.  In this paper, two types of item embedding are used and learned: one is the typical item embedding, and the other one is graph-based embedding. For the typical item embedding, we project all items into \(M_s \in \mathbb{R}^{|I| \times d}\) via an embedding layer, where \(|I|\) denotes the total number of items, and \(d\) represents the dimension of the embedding. For the graph-based item embedding, we use a GCN-based graph encoder to project all items into an embedding space.
\subsection{Graph convolutional encoder}
\label{sec:graphencoder}
In particular, for the GCN-based graph encoder, we draw upon the concepts presented in \cite{chen2020revisiting, 10.1145/3397271.3401063}. We construct a single graph for the entire dataset to capture node relationships from a global perspective. To build the graph for items, each item within a dataset is viewed as a node.  If two items are co-located in less than $z$ distance in a historical sequence, we add an edge between them. Here, $z$ represents a predetermined maximum distance. Initially, we project all items into a common embedding space, \(M_g^0 \in \mathbb{R}^{|I| \times d}\), where $|I|$ is the number of items and $d$ is the dimension of embedding, and we treat this as the first layer's item embedding in the graph. Following \cite{10.1145/3397271.3401063}, we discard feature transformation and nonlinear activation for improving efficiency. 
Then the computation within the GCN-based graph encoder proceeds as follows:
\begin{equation}
    \begin{aligned}
    \label{eq:3}        \mathbf{m}_i^{l+1}=\mathbf{m}_i^l+\sum_{i^{\prime} \in \mathcal{N}_i} \mathbf{m}_{i^{\prime}}^l ; \quad \tilde{\mathbf{m}}_i=\sum_{l=0}^L \mathbf{m}_i^l
    \end{aligned}
\end{equation}
where $L$ denotes the total number of layers, and $\mathcal{N}_i$ represents one-hop neighbor nodes of $m_i$. $\mathbf{m}_i^l, \mathbf{m}_{i^{\prime}}^l$ represent the embedding of items $i, i^{\prime} \in |I|$ in the $l$-th layer. Specifically, we sum up the representations from all layers to obtain the final embedding of an item $i$, denoted as $\tilde{\mathbf{m}}_i$. We call it graph-based (item) embedding, and all items' graph-based embeddings are represented as a matrix \(M_g \in \mathbb{R}^{|I| \times d}\). The graph encoder is designed to convert items into expressive representations based on the structural information in the graph.
\subsection{Transformer-based sequence encoder}
\label{sec:encoder}
Transformer-based sequence encoder is a vital step in the sequential recommendation. It aims to extract the representation from the sequence list. First of all, we describe input to the sequence encoder. 

\smallskip
\noindent\textbf{Input to the sequence encoder.} 
Given the input as an interaction history sequence \(S^u = [c_1, c_2, ..., c_n]\), the Transformer takes into account the positions of items by initializing the history item list \(S^u\) to \(e^u \in \mathbb{R}^{n \times d}\) by:
\begin{equation}
\begin{aligned}
e_s^u = [m_{s_1} + p_1, m_{s_2} + p_2, ..., m_{s_n} + p_n].\\
e_g^u = [m_{g_1} + p_1, m_{g_2} + p_2, ..., m_{g_n} + p_n].
\end{aligned}
\end{equation}
where \(m_{s_i} \in \mathbb{R}^d\) represents an item's typical item embedding at the \(i\)-th position in the sequence, \(m_{g_i} \in \mathbb{R}^d\) represents the item's graph-based embedding at \(i\)-th position in the sequence, \(p_i \in \mathbb{R}^d\) denotes the positional embedding, and \(n\) is the sequence length. We note that $m_{s_i}$ and $m_{g_i}$ are extracted from embedding matrices $M_s$ and $M_g$, respectively, described in the previous subsections.

\smallskip
\noindent\textbf{Sequence encoder.} The sequence encoder derives the representation of $e^u$ using a deep neural network (e.g., BERT4Rec) \cite{sun2019bert4rec}. We use two sequence encoders: one for the sequence of item-based embeddings ($e_s^u$) and the other one for the same sequence of graph-based embeddings ($e_g^u$). The sequence encoders are defined as $f_\theta$ and $f_\phi$, respectively, where $\theta$ and $\phi$ represent each model's parameters. The output representation $H_s^u \in R^{n\times d}$ and $H_g^u \in R^{n\times d}$ are calculated as follows:
\begin{equation}
\begin{aligned}
\label{eq:1}
    H_s^u = f_\theta(e_s^u)\\
    H_g^u = f_\phi(e_g^u)
\end{aligned}
\end{equation}
Since our main task is to predict the next item, we employ the final vectors $h_{s_n}$ in $H_s^u=[h_{s_1}, h_{s_2}, ..., h_{s_n}]$ and $h_{g_n}$ in $H_g^u=[h_{g_1}, h_{g_2}, ..., h_{g_n}]$ as the item-based sequence representation and graph-based sequence representation of the historical sequence, respectively. We can interpret them as two types of user representation.

\subsection{Multi-view contrastive learning}
\smallskip
\noindent\textbf{Inner view contrastive learning.}
Inspired by the prior work \cite{xie2022contrastive,Qin_2023}, we utilize InfoNCE as the objective function to optimize features extracted from contrastive learning. We denote the number of historical sequences in each batch by \textit{B}. Given $B$ historical sequences in the batch, each historical sequence goes through the stochastic data augmentation module and returns two augmented sequences, so totally there are \textit{2B} augmented sequences. Since contrastive learning requires positive pairs and negative pairs, given a user's historical sequence (i.e., one of \textit{B} historical sequences in the batch), we create a positive pair of the sequence via the stochastic data augmentation module. We use the remaining $2(B-1)$ augmented sequences as negative samples for the positive pair. 

For each positive pair, contrastive loss is calculated by:

\begin{equation}
\begin{aligned}
\mathcal{L}_{\text {con }}\left(h_n^1, h_n^2\right)= 
& -\log \frac{exp^{s\left(h_n^1, h_n^2\right)}}{exp^{s\left(h_n^1, h_n^2\right)}+\sum_{h_n \in \text { neg }} exp^{s\left(h_n^1, h_n\right)}} \\
& -\log \frac{exp^{s\left(h_n^2, h_n^1\right)}}{exp^{s\left(h_n^2, h_n^1\right)}+\sum_{h_n \in \text { neg }} exp^{s\left(h_n^2, h_n\right)}}
\end{aligned}
\end{equation}
where $h_n^1$ and $h_n^2$ are the positive pair's sequence representations learned from the same Transformer-based sequence encoder (i.e., either $f_\theta$ or $f_\phi$). $s(,)$ represents the inner product, and $neg$ indicates the set of negative sample embeddings/representations. Since we can create $2(B-1)$ negative pairs for each of $h_n^1$ and $h_n^2$, the loss function consists of two terms. 

Then, the objective function for optimizing the contrastive learning over the two different views (i.e., item-based sequence representation and graph-based sequence representation via the sequence encoders) is as follows:

\begin{equation}
\begin{aligned}
\label{eq:5}
\mathcal{L}_{\text {Inner}} = \mathcal{L}_{\text {con }}\left(h_{s_n}^1, h_{s_n}^2\right) + \mathcal{L}_{\text {con }}\left(h_{g_n}^1, h_{g_n}^2\right)
\end{aligned}
\end{equation}
Where $h_{s_n}^1$ and $h_{s_n}^2$ are item-based sequence representations of the positive pair, and $h_{g_n}^1$ and $h_{g_n}^2$ are graph-based sequence representations of the positive pair.

\smallskip
\noindent\textbf{Cross-view contrastive learning.}
In addition to the inner view contrastive learning, we propose a cross view contrastive learning, which learns discriminative features that capture the correspondence between the item-based sequence representation and graph-based sequence representation. 

Firstly, the mean of the $h_{s_n}^1$ and $h_{s_n}^2$ obtained from a positive pair is calculated as $h_{s_n}^m$, and the mean of the $h_{g_n}^1$ and $h_{g_n}^2$ obtained from the same positive pair is calculated as $h_{g_n}^m$. Likewise, given \textit{2(B-1)} negative samples in the batch, each two negative samples were originated from the same historical sequence (i.e., \textit{B-1} negative sample pairs). Therefore, we also get each negative sample pair's mean of item-based sequence representations and mean of graph-based sequence representations. 

Given the positive pair's mean representations ${h_{s_n}^m}$ and ${h_{g_n}^m}$, cross-view contrastive loss is calculated as follows:
\begin{equation}
\begin{aligned}
\label{eq:6}
\mathcal{L}_{\text {cross}}= \mathcal{L}_{\text {con }}\left(h_{s_n}^m, h_{g_n}^m\right)
\end{aligned}
\end{equation}
$\mathcal{L}_{\text{cross}}$ is designed to maximize the similarity between $h_{g_n}^m$ and $h_{s_n}^m$. 
This approach compels the model to learn similar item-based and graph-based representations of the same historical sequence (or augmented sequences originated from the same sequence), yielding enhanced representation capability.

Consequently, we combine the aforementioned two contrastive loss functions as follows:
\begin{equation}
\begin{aligned}
\label{eq:mm}
\mathcal{L}_{\text{MM}} = \mathcal{L}_{\text {cross}} + \mathcal{L}_{\text {Inner}}
\end{aligned}
\end{equation}

\subsection{Multi-view attention fusion module \& recommendation prediction}
To further utilize the extracted representations, we propose a multi-view attention fusion module designed to amalgamate information from two disparate views -- item-based sequences and graph-based sequences -- and ultimately predict a target user's preference score for a given item.

The multi-view attention fusion module is executed through an interactive cross-view attention mechanism, which is devised to uncover multi-view global and local dependencies. Given a user's two different view representations, \( h_{g_n}^m \in R^{1\times d}\) and \( h_{s_n}^m \in R^{1\times d}\), as depicted in Figure~\ref{fig:2} (the rightmost figure), we initially calculate the global attention score, \( s_{\text{global}}^{\text{attention}} \), and the local attention score, \( s_{\text{local}}^{\text{attention}} \):
\begin{equation}
    \begin{aligned}
    \label{eq:7}
        & s_{global}^{attention} = Averagepool\left(h_{g_n}^m+h_{s_n}^m\right)\otimes W_g\\
        & s_{local}^{attention} = \sigma\left(\left(h_{g_n}^m+h_{s_n}^m\right)\otimes W_l\right)\\
    \end{aligned}
\end{equation}
where \( W_g \in R^{1\times 1}\) and \( W_l \in R^{d\times d}\) represent global weight and local weight matrix, respectively. \textit{Averagepool} represents an average pool function, which deals the representations from a global view, returns a single numeric value, which is the average of the $d$ values. $d$ is the dimension of a view's sequence representation, and \( \otimes \) denotes matrix product. \( \sigma \) represents an activation function. In this paper, we employ ReLU as the activation function. 


Given the global and local attention scores, a new fusion task arises as follows:
\begin{equation}
    \begin{aligned}
    \label{eq:8}
        s^{attention} = \text{sigmoid} \left(s_{global}^{attention}\oplus s_{local}^{attention}\right)
    \end{aligned}
\end{equation}
where \( \oplus \) represents the summation between the $s_{global}^{attention}$ and $s_{local}^{attention}$ with broadcasting to handle their different dimensions. We employ the sigmoid function to normalize the scores. These scores are considered as weights for different view representations. 

Finally, fused representation $f_{fusion}\left(h_{s_n}^m, h_{g_n}^m\right)\in R^{1\times d}$is calculated as follows:
\begin{equation}
    \begin{aligned}
    \label{eq:9}
        &f_{fusion}\left(h_{s_n}^m, h_{g_n}^m\right) \\&=  \left(s^{attention}\circ h_{g_n}^m\right)\oplus \left(\left(1-s^{attention}\right)\circ h_{s_n}^m\right)
    \end{aligned}
\end{equation}
where $\circ$ represents elements-wise product. 

Since our final goal is to use this representation for recommendation, we propose a novel strategy to leverage the generated representation:
\begin{equation}
    \begin{aligned}
    \label{eq:ss}
        \hat{y} =  f_{fusion}\left(h_{s_n}^m, h_{g_n}^m\right)M_s^T+f_{fusion}\left(h_{g_n}^m, h_{s_n}^m\right)M_g^T
    \end{aligned}
\end{equation}
where $h_{s_n}^m$ and $h_{g_n}^m$ are a target user's item-based sequence representation and graph-based sequence representation, respectively. $M_s$ and $M_g$ are the typical item embedding matrix and graph-based item embedding matrix, respectively, described in Section~\ref{sec:embeddings}. 

In our paper, we utilize cross-entropy loss as the objective function, optimizing to improve prediction accuracy.
\begin{equation}
    \begin{aligned}
    \label{eq:10}
        \mathcal{L}_{rec} =  H(y, \hat{y}) = - \sum_{i} y_i \log(\hat{y}_i)
    \end{aligned}
\end{equation}
where $y$ represents the ground truth label for the user's true preference scores to items.

\subsection{Overall Objective}
Finally, the total loss function during the training stage can be represented as:
\begin{equation}
    \begin{aligned}
    \label{eq:14}
        \mathcal{L} =  \mathcal{L}_{rec} + \lambda\mathcal{L}_{MM}
    \end{aligned}
\end{equation}
where $\mathcal{L}_{rec}$ is the recommendation objective function in Eq. \ref{eq:10}, $\mathcal{L}_{MM}$ represents multi-view contrastive loss, which consists of the inner-view contrastive loss and cross-view contrastive loss, as defined in Eq. \ref{eq:mm}, and $\lambda$ is a hyperparameter.

\section{Time Complexity}
The time complexity of MVCRec is primarily influenced by the training phase, which includes both multi-view contrastive learning and embedding calculations. Training is split into two stages. In the first stage, MVCRec generates augmented views of user sequences and computes item-based and graph-based embeddings, resulting in a complexity of \( O(|U|d^2 + |U|d) \), where \( |U| \) is the number of users and \( d \) is the embedding dimension.

The second stage, which dominates the complexity, involves inner-view and cross-view contrastive learning with a complexity of \( O(|U|^2d) \). 
This is due to pairwise comparisons across augmented views within each type (item-based and graph-based), enhancing MVCRec’s ability to capture user preferences.

However, in the testing phase, the model only requires the multi-view attention fusion module, simplifying the complexity to \( O(nd) \), similar to that of an attention-based encoder like SASRec, where \( n \) is the sequence length. 
\section{Experiment}
In this section, we conduct extensive experiments using five real-world datasets to investigate the following research questions (RQs):

\squishlist
\item \textbf{RQ1}: How is the performance of our MVCrec compared with existing baselines?

\item \textbf{RQ2}: How effective are the key components of MVCrec in terms of enhancing the model's performance?


\item \textbf{RQ3}: How do hyperparameters (i.e., the weight of the multi-view contrative loss $\lambda$, a batch size, and an embbeding size) affect the performance of MVCrec?
\squishend

\subsection{Experimental settings}
\subsubsection{\textbf{Dataset}}
To verify the effectiveness of our model, we evaluate its performance using five real-world benchmark datasets: Amazon (Beauty, Sports and Home \& Kitchen)\footnote{https://jmcauley.ucsd.edu/data/amazon/}, Yelp\footnote{https://www.yelp.com/dataset} and Reddit dataset\footnote{https://www.kaggle.com/datasets/colemaclean/subReddit-interactions/data}. The Amazon datasets contain a series of Amazon product reviews. In our experiments, we use three sub-categories of the Amazon: Beauty, Sports, and Home \& Kitchen. The Yelp dataset, containing reviews of businesses listed on Yelp, serves a similar purpose as the Amazon datasets. Reddit dataset contains user interaction on the Reddit platform. 
In the following experiments, we only use interaction data without any auxiliary data (e.g., text, image). Following the preprocessing steps described in \cite{liu2021contrastive, wang2023sequential}, we removed users and items with fewer than five interactions. The statistics of the datasets are summarized in Table~\ref{tabl1}.
\begin{table}[ht]
	\centering
	\caption{The statistics of datasets.}
    \resizebox{\linewidth}{!}{
	\begin{tabular}{cccccc}
		\hline
		Dataset & \#users & \#items & \#interactions & avg.length & sparsity\\
        \hline
        Sports & 33.6K & 18.3K & 296.3K & 8.3 & 99.95\% \\
        Beauty & 22.3K & 12.1K & 198.5K & 8.8 & 99.93\% \\
        Yelp & 30.4K & 20.0K & 316.3K & 10.4 & 99.95\% \\
        Home \& Kitchen & 66.5K & 28.2K & 551.6K & 8.3 & 99.97\% \\
        Reddit & 14.5K & 15.4K & 28.9K & 20.95 & 99.99\% \\
        \hline
	\end{tabular}%
    }
	\label{tabl1}%
\end{table}%
\subsubsection{\textbf{Baselines}} We compare our model with 11 state-of-the-art recommendation models, which can be divided into three parts:

\smallskip
\noindent\textbf{Non-sequential models.} These baselines are based on collaborative learning and graph convolutional network:
\squishlist
\item \textit{BPRMF} \cite{10.5555/1795114.1795167} uses Bayesian Personalized Ranking (BPR) loss to optimize the matrix factorization model.
\item \textit{LightGCN} \cite{10.1145/3397271.3401063} simplifies the design of GCN to make it more concise and appropriate for recommendation.
\squishend 

\smallskip
\noindent\textbf{General sequential models.} These baselines are based on RNN, attention-based neural networks, memory neural networks, GCN-based networks:
\squishlist

\item \textit{SRGNN} \cite{Wu_2019} models the historical item sequence as a graph-structured data to deal with sequential recommendation.
\item \textit{GRU4rec} \cite{DBLP:journals/corr/HidasiKBT15} uses Gated Recurrent Unit (GRU) to model for the sequential recommendation.
\item \textit{Caser} \cite{tang2018personalized} embeds a sequence of recent items into an “image” in the time and latent spaces, and learns sequential patterns as local features of the image using convolutional filters.
\item \textit{SASRec} \cite{kang2018self} proposes the first self-attention based sequential model to capture long-term dependencies.
\squishend

\smallskip
\noindent\textbf{Self-supervised sequential models.} These baselines are based on Transformer and collaborative learning:
\squishlist
\item \textit{BERT4Rec} \cite{sun2019bert4rec} trains the bidirectional model using the Cloze task, predicting the masked items in the sequence by jointly conditioning on their left and right context.
\item \textit{CL4rec} \cite{xie2022contrastive} leverages contrastive learning on the sequential recommendation.
\item \textit{MCLSR} \cite{wang2022multi} learns the representations of users and items through a cross-view contrastive learning paradigm from four specific views at two different levels (i.e., interest- and feature-level).
\item \textit{MCLrec} \cite{Qin_2023} innovates the standard contrastive learning framework by contrasting data, and models augmented views for adaptively capturing the informative features hidden in stochastic data augmentation. 
\item \textit{DCrec} \cite{yang2023debiased} 
proposes a global collaborative learning strategy to tackle with the popularity bias for sequential recommendation, considering dependencies between users across sequences.
\squishend

We note that other existing methods (e.g., UltraGCN, VGCL, CGCL, MAErec, MMSSL \cite{wei2023multi}, MSM4SR \cite{zhang2023multimodal}) which are not aimed for sequential recommendation or require auxiliary data, are excluded in the baseline list except well-known \textit{BPRMF} and \textit{LightGCN} because their performance would be much lower than sequential recommendation models or sometimes it is hard to run some of their models without auxiliary data (again, in this paper, we only utilize a target behavior's interaction data without any auxiliary data).
\begin{table*}
	\centering
	\caption{Overall coverage where bold means the best performance and underline means the second-best performance. The p-value for the result is less than 0.01.}
    \resizebox{\linewidth}{!}{
	\begin{tabular}{c|c|ccccccccccccc}
		\hline		Dataset & Metric & BPRMF & LightGCN & GRU4rec & Caser & SASRec & BERT4Rec & SRGNN & CL4rec & MCLrec & MCLSR & DCrec & MVCrec & Improv.(\%)\\
\hline
Sport & HR@5 & 0.0144 & 0.0171 & 0.0113 & 0.0060 & 0.0242 & 0.0222 & 0.0214 & 0.0258 & 0.0281 & 0.0186 & \underline{0.0333} & \textbf{0.0352} & 5.71\\
~ & NDCG@5 & 0.0092 & 0.0107 & 0.0073 & 0.0043 & 0.0158 & 0.0147 & 0.0144 & 0.0171 & 0.0191 & 0.0123 & \underline{0.0231} & \textbf{0.0238} & 3.03\\
~ & HR@10 & 0.0255 & 0.0289 & 0.0182 & 0.0092 & 0.0369 & 0.0351 & 0.0330 & 0.0403 & 0.0428 & 0.0287 & \underline{0.0481} & \textbf{0.0523} & 8.73\\
~ & NDCG@10 & 0.0127 & 0.0146 & 0.0095 & 0.0053 & 0.0199 & 0.0189 & 0.0181 & 0.0218 & 0.0239 & 0.0155 & \underline{0.0278} & \textbf{0.0293} & 5.4\\
~ & HR@20 & 0.0414 & 0.0471 & 0.0317 & 0.0138 & 0.0550 & 0.0527 & 0.0508 & 0.0607 & 0.0662 & 0.0444 & \underline{0.0683} & \textbf{0.0760} & 11.27\\
~ & NDCG@20 & 0.0168 & 0.0191 & 0.0129 & 0.0065 & 0.0245 & 0.0233 & 0.0226 & 0.0269 & 0.0297 & 0.0195 & \underline{0.0329} & \textbf{0.0352} & 6.99\\

\hline
Beauty & HR@5 & 0.0235 & 0.0262 & 0.0166 & 0.0107 & 0.0466 & 0.0439 & 0.0433 & 0.0516 & 0.0564 & 0.0275 & \underline{0.0614} & \textbf{0.0647} & 5.37\\
~ & NDCG@5 & 0.0143 & 0.0165 & 0.0108 & 0.0068 & 0.0311 & 0.0291 & 0.0304 & 0.0354 & 0.0388 & 0.0189 & \underline{0.0439} & \textbf{0.0460} & 4.78\\
~ & HR@10 & 0.0397 & 0.0433 & 0.0273 & 0.0174 & 0.0656 & 0.0643 & 0.0620 & 0.0749 & 0.0837 & 0.0410 & \underline{0.0846} & \textbf{0.0924} & 9.22\\
~ & NDCG@10 & 0.0195 & 0.0220 & 0.0142 & 0.0089 & 0.0372 & 0.0356 & 0.0364 & 0.0428 & 0.0476 & 0.0233 & \underline{0.0513} & \textbf{0.0548} & 6.82\\
~ & HR@20 & 0.0614 & 0.0695 & 0.0446 & 0.0267 & 0.0944 & 0.0935 & 0.0910 & 0.1068 & 0.1166 & 0.0639 & \underline{0.1145} & \textbf{0.1275} & 11.35\\
~ & NDCG@20 & 0.0250 & 0.0286 & 0.0186 & 0.0113 & 0.0444 & 0.0430 & 0.0437 & 0.0509 & 0.0560 & 0.0290 & \underline{0.0588} & \textbf{0.0637} & 8.33\\

\hline
Yelp & HR@5 & 0.0336 & 0.0502 & 0.0134 & 0.0060 & 0.0409 & 0.0419 & 0.0269 & 0.0447 & \underline{0.0531} & 0.0491 & 0.0478 & \textbf{0.0597} & 12.43\\
~ & NDCG@5 & 0.0223 & 0.0357 & 0.0082 & 0.0043 & 0.0331 & 0.0337 & 0.0180 & 0.0328 & \underline{0.0380} & 0.0342 & 0.0374 & \textbf{0.0447} & 17.63\\
~ & HR@10 & 0.0512 & 0.0730 & 0.0218 & 0.0092 & 0.0551 & 0.0562 & 0.0431 & 0.0642 & \underline{0.0751} & 0.0714 & 0.0654 & \textbf{0.0811} & 7.99\\
~ & NDCG@10 & 0.0280 & 0.0430 & 0.0109 & 0.0053 & 0.0377 & 0.0383 & 0.0232 & 0.0391 & \underline{0.0450} & 0.0414 & 0.0431 & \textbf{0.0515} & 14.44\\
~ & HR@20 & 0.0812 & 0.1060 & 0.0371 & 0.0138 & 0.0778 & 0.0800 & 0.0673 & 0.0938 & \underline{0.1076} & 0.1025 & 0.0913 & \textbf{0.1107} & 2.88\\
~ & NDCG@20 & 0.0355 & 0.0513 & 0.0147 & 0.0065 & 0.0434 & 0.0443 & 0.0293 & 0.0466 & \underline{0.0532} & 0.0492 & 0.0496 & \textbf{0.0589} & 10.71\\

\hline
Home \& kitchen & HR@5 & 0.0054 & 0.0073 & 0.0039 & 0.0042 & 0.0113 & 0.0116 & 0.0066 & 0.0141 & 0.0153 & 0.0046 & \underline{0.0198} & \textbf{0.0207} & 4.55\\
~ & NDCG@5 & 0.0035 & 0.0046 & 0.0024 & 0.0026 & 0.0074 & 0.0076 & 0.004 & 0.0096 & 0.0106 & 0.0029 & \underline{0.0146} & \textbf{0.0147} & 0.68\\
~ & HR@10 & 0.0094 & 0.0122 & 0.0066 & 0.0072 & 0.0180 & 0.0172 & 0.0116 & 0.0211 & 0.0227 & 0.0079 & \underline{0.0269} & \textbf{0.0288} & 7.06\\
~ & NDCG@10 & 0.0048 & 0.0061 & 0.0032 & 0.0036 & 0.0096 & 0.0094 & 0.0057 & 0.0118 & 0.0129 & 0.0039 & \underline{0.0169} & \textbf{0.0171} & 1.18\\
~ & HR@20 & 0.0158 & 0.0202 & 0.0127 & 0.0129 & 0.0275 & 0.0265 & 0.0196 & 0.0307 & 0.0331 & 0.0144 & \underline{0.0362} & \textbf{0.0407} & 12.43\\
~ & NDCG@20 & 0.0064 & 0.0082 & 0.0048 & 0.0050 & 0.0120 & 0.0117 & 0.0077 & 0.0142 & 0.0156 & 0.0056 & \underline{0.0193} & \textbf{0.0201} & 4.15\\

\hline
Reddit & HR@5 & 0.2805 & 0.3103 & 0.1820 & 0.1662 & 0.2132 & 0.2221 & 0.2458 & 0.3072 & \underline{0.3217} & 0.1767 & 0.3142 & \textbf{0.3374} & 4.88\\
~ & NDCG@5 & 0.2241 & 0.2444 & 0.1525 & 0.1418 & 0.1726 & 0.1756 & 0.1881 & 0.2436 & \underline{0.2557} & 0.1459 & 0.2477 & \textbf{0.2661} & 4.07\\
~ & HR@10 & 0.3418 & 0.3744 & 0.2166 & 0.1920 & 0.2618 & 0.2750 & 0.3024 & 0.3769 & \underline{0.3859} & 0.2117 & 0.3727 & \textbf{0.4053} & 5.03\\
~ & NDCG@10 & 0.2438 & 0.2650 & 0.1636 & 0.1501 & 0.1883 & 0.1928 & 0.2064 & 0.2561 & \underline{0.2765} & 0.1571 & 0.2667 & \textbf{0.2880} & 4.16\\
~ & HR@20 & 0.4098 & 0.4526 & 0.2626 & 0.2267 & 0.3280 & 0.3444 & 0.3655 & 0.4483 & \underline{0.4560} & 0.2615 & 0.4417 & \textbf{0.4773} & 4.67\\
~ & NDCG@20 & 0.2610 & 0.2847 & 0.1752 & 0.1588 & 0.2050 & 0.2103 & 0.2223 & 0.2741 & \underline{0.2941} & 0.1696 & 0.2840 & \textbf{0.3063} & 4.15\\

\hline
	\end{tabular}%
    }
	\label{tab:overall-performance}%
\end{table*}
\subsubsection{\textbf{Evaluation metric}}
In accordance with \cite{10.1145/3543507.3583529, 10.1145/3543507.3583479, xie2022contrastive, huang2023modeling, he2022query}, we employ the leave-one-out strategy to split each dataset into training, validation, and test sets based on the timestamp provided by the dataset. Specifically, we use the last interaction of every user for the test set, and the second-to-last interaction for every user is allocated for the validation set; all remaining interactions are used in the training set. Following the procedure in \cite{hou2023learning, fan2022sequential, lin2022sequential, li2023exploiting}, we rank the entire item set.

We adopt Hit Ratio (HR) and Normalized Discounted Cumulative Gain (NDCG) as evaluation metrics. HR@k measures whether the positive item appears in the top-k recommendation list, and NDCG@k additionally considers its position in the ranking list, where $k \in \{5, 10, 20\}$.

\subsubsection{\textbf{Implementation Details}}
We implement our method using PyTorch, aligning the implementation of BPRMF, LightGCN, FPMC, GRU4rec, Case, SASRec, CL4rec, and BERT4Rec with the methodologies described in their respective papers. 
We implemented MCLSR ourselves as the authors did not provide their code in the paper. A graph for the graph encoder is constructed based on the training set. To ensure fairness, we employ BERT as the representation encoder for CL4rec, MCLrec, DCrec, and our MVCrec, setting the number of self-attention blocks and attention heads to 2, and we set the item-to-item distance $z$ as 3 (as mentioned in Section \ref{sec:graphencoder}).
All parameters are consistent with those reported in the original papers, and optimal settings are chosen based on model performance on the validation set. We set the embedding size \(d\) as 64 and the maximum length of recently consumed items in each user's historical sequence \(n\) as 50, selecting a hyperparameter \(\lambda\) from \(\{0.01, 0.02, 0.03, 0.04, 0.05, 0.1, 0.2, 0.3, 0.4, 0.5\}\). The learning rate \(lr\) is chosen from \(\{1e-3, 1e-4\}\), and weight decay is selected from \(\{0,1e-1,1e-2,1e-3,1e-4,1e-5,1e-6\}\). For fairness, we standardize the batch size \textit{B} to 256 for all models and fix temperature of contrastive learing for all contrastive learning method as 1. 
The models are optimized using the Adam optimizer \cite{10.1145/3539618.3591691} and are trained with an early stopping strategy based on the performance of the validation set, with the maximum step set to 100. All experiments are conducted on a Tesla T4 GPU.

\begin{figure*}[th]
     \centering
     \begin{subfigure}[b]{0.40\textwidth}
         \centering
         \includegraphics[width=\textwidth]{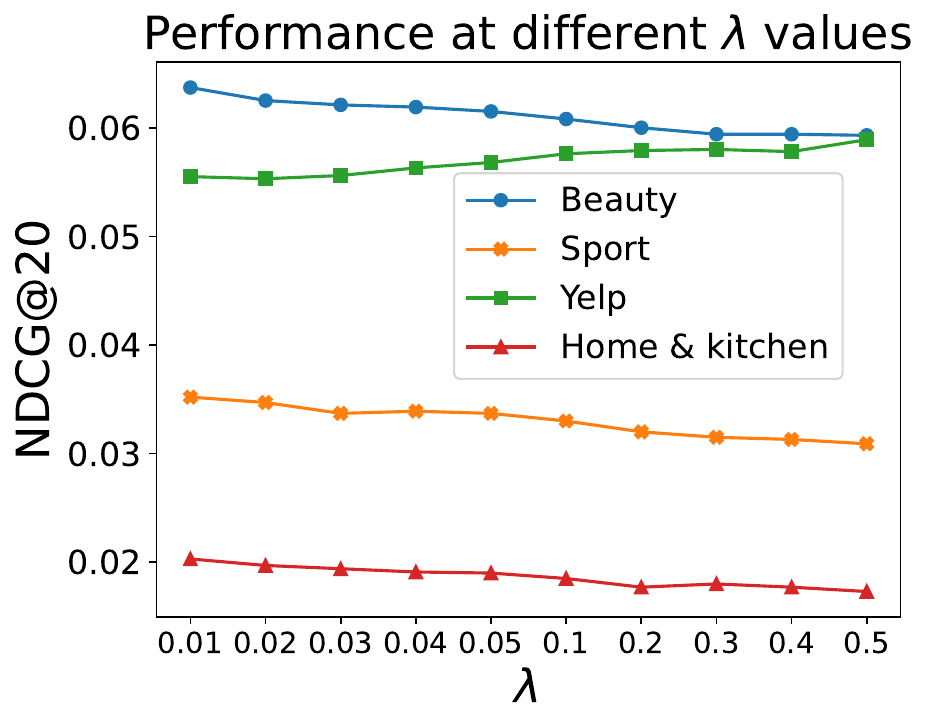}
         \caption{Amazon and Yelp}
         \label{fig:lambda_amazon_yelp}
     \end{subfigure}
     \begin{subfigure}[b]{0.40\textwidth}
         \centering
         \includegraphics[width=\textwidth]{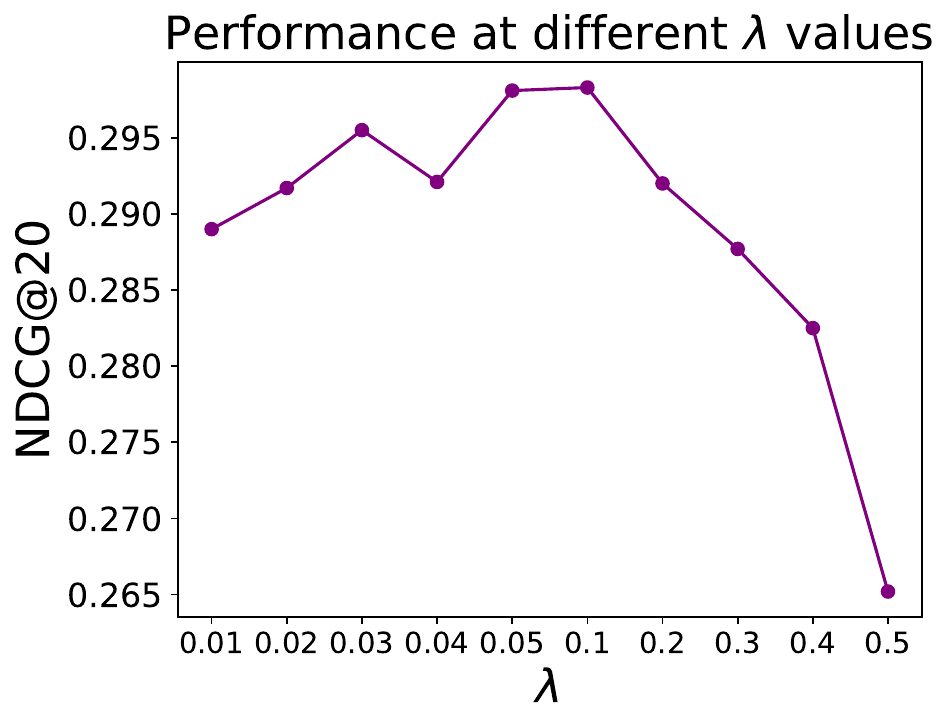}
         \caption{Reddit}
         \label{fig:lambda_reddit}
     \end{subfigure}
    \caption{Performance at different $\lambda$ under NDCG@20.}
    \label{fig:lambda}
\end{figure*}

\subsection{RQ1: overall performance}
To clarify the contributions made by MVCrec, we compare its performance with the baselines. The results presented in Table~\ref{tab:overall-performance} lead us to several insights:

\squishlist
\item Our method outperforms the baselines, attributed to the graph view and the multi-view fusion strategy. For instance, our model surpasses the best baseline by 1.18\%$\sim$14.44\% on NDCG@10 and 5.03\% $\sim$ 9.22\% on HR@10 over the five datasets. This superior performance can be explained as follows: (1) The multi-view contrastive learning strategy incorporating both sequence and graph information facilitates the generation of more expressive representations; and (2) The multi-view attention fusion strategy effectively amalgamates item-based sequence representation and graph-based sequence representation. These results confirm the effectiveness of our multi-view contrastive recommendation model, learning more accurate and better representations.

\item Self-supervised models (e.g., MCLrec, DCrec) exhibit pronounced efficacy, markedly surpassing classical models such as BPRMF, LightGCN, GRU4rec and Caser. Fundamental Transformer-based methods like SASRec and BERT4Rec excel beyond the classical models, establishing themselves as the secondary tier in sequential recommendation and emphasizing the power of Transformer methods in this realm. In contrast to SASRec and BERT4Rec, models like CL4rec, MCLrec, DCrec, and MVCrec integrate contrastive learning and data augmentation methods for training in the recommendation tasks. This demonstrates contrastive learning's capability to harness more intricate representations from historical sequences by learning features that discern between distinct instances. Interestingly, LightGCN manifests substantial prowess on the Yelp dataset, aligning closely with CL4rec and underscoring the proficiency of graph networks in recommendation systems.

\item In comparison to CL4rec, our findings substantiate that a graph structure tailored for sequential recommendation can notably enhance performance. LightGCN also eclipses BPRMF substantially, elevating the graph structure; the graph convolutional network unveils connections between users and items as their interaction is inherently graphical. Concurrently, the results show that our sequence-based graph construction method adeptly discern interactions between varied items by weighing the positioning of items within the sequence.
\squishend


\subsection{RQ2: ablation study}
Next, we conduct an ablation study to test whether each proposed component positively contribute to the performance improvement or not. 

\begin{table}[ht]
	\centering
	\caption{Ablation study at HR@20 and NDCG@20.}
 \resizebox{\linewidth}{!}{
	\begin{tabular}{c|c|c|c|c|c}
        \hline
        \multicolumn{2}{c|}{Model} & MVCrec & MVCrec(s) & MVCrec(g) & MVCrec(mlp)\\
        \hline
        \multirow{2}*{Beauty} & HR & 0.1275 & 0.1068 & 0.1183 & 0.1017\\
        ~ & NDCG & 0.0637 & 0.0509 & 0.0559 & 0.0489\\
        \hline
        \multirow{2}*{Sport} & HR & 0.0760 & 0.0607 & 0.0705 & 0.0590\\
        ~ & NDCG & 0.0352 & 0.0269 & 0.0319 & 0.0271\\
        \hline
        \multirow{2}*{Yelp} & HR & 0.1107 & 0.0938 & 0.1020 & 0.0923\\
        ~ & NDCG & 0.0589 & 0.0466 & 0.0523 & 0.0449\\
        \hline
        \multirow{2}*{Home \& Kitchen} & HR & 0.0407 & 0.0307 & 0.0362 & 0.0349\\
        ~ & NDCG & 0.0201 & 0.0142 & 0.0164 & 0.0160\\
        \hline
        \multirow{2}*{Reddit} & HR & 0.4773 & 0.4218 & 0.4348 & 0.4013\\
        ~ & NDCG & 0.3063 & 0.2697 & 0.2780 & 0.2651\\
        \hline
	\end{tabular}%
 }
	\label{tb3}%
\end{table}

To further comprehend the efficacy of our proposed model MVCrec, we compare it with three variants of our model: MVCrec(s), MVCrec(g) and MVCrec(mlp). MVCrec(s) employs a single contrastive learning approach based on \textit{only item-based sequence information} without graph-based sequence information. MVCrec(g) denotes utilization of contrastive learning \textit{solely on the graph-based sequence information} without item-based sequence information. MVCrec(mlp) denotes the use of multilayer perceptron (MLP) by concatenating representations of two views as input and then going through the MLP layers instead of our proposed multi-view attention fusion module. Following \cite{Qin_2023}, we adopt HR@20 and NDCG@20 as evaluation metrics in the ablation study for simplification. 

The results are presented in Table \ref{tb3}. Analyzing the comparison between our model and three variants yields the following insights: 

\squishlist
\item A comparison between MVCrec(s) and MVCrec(g) reveals that the graph convolutional layer is more pivotal in terms of representing the history sequence. The information in the graph, constructed by the sequence data, encapsulates extensive user preference. 

\item Comparing MVCrec(s) and MVCrec shows that our full method significantly outperforms MVCrec(s) -- analogous to CL4rec -- attributed to our novel multi-view attention fusion module that harnesses information from both graph and sequence structures to generate more expressive representations. 

\item Comparing MVCrec(mlp) and MVCrec shows that our proposed multi-view attention fusion module outperforms the MLP significantly, this is because the multi-view attention fusion module utilizes attention to weigh the importance of different views and their features dynamically.
\squishend

\subsection{RQ3: hyperparameter analysis}

\subsubsection{Hyperparameter Analysis on $\lambda$}\par
In this section, we examine the impact of varying $\lambda$, a hyperparameter in Eq.~\ref{eq:14}. We assess the performance of MVCrec across five datasets using different values of $\lambda$. Because of the limited space, we report NDCG@20 as the evaluation metric, and the results are illustrated in Figure \ref{fig:lambda}. In the Amazon datasets (i.e., Beauty, Sports, and Home \& Kitchen), optimal performance is achieved when $\lambda$ is set to 0.01. In the Yelp dataset, performance improves with increasing $\lambda$, reaching its highest at 0.5. For the Reddit dataset, the best performance is observed with $\lambda$ set to 0.1. It means both recommendation loss and contrastive loss positively contributed to correctly estimate user-item matching scores and learn better representations. The discrepancy of optimal $\lambda$ among the datasets can be potentially explained that Amazon dataset have smaller average historical sequence length than Yelp and Reddit datasets. Although we do not report HR@20, We observed similar trends in both HR@20 and NDCG@20.
\begin{figure}[htb]
  \centering
  \includegraphics[width=0.85\linewidth]{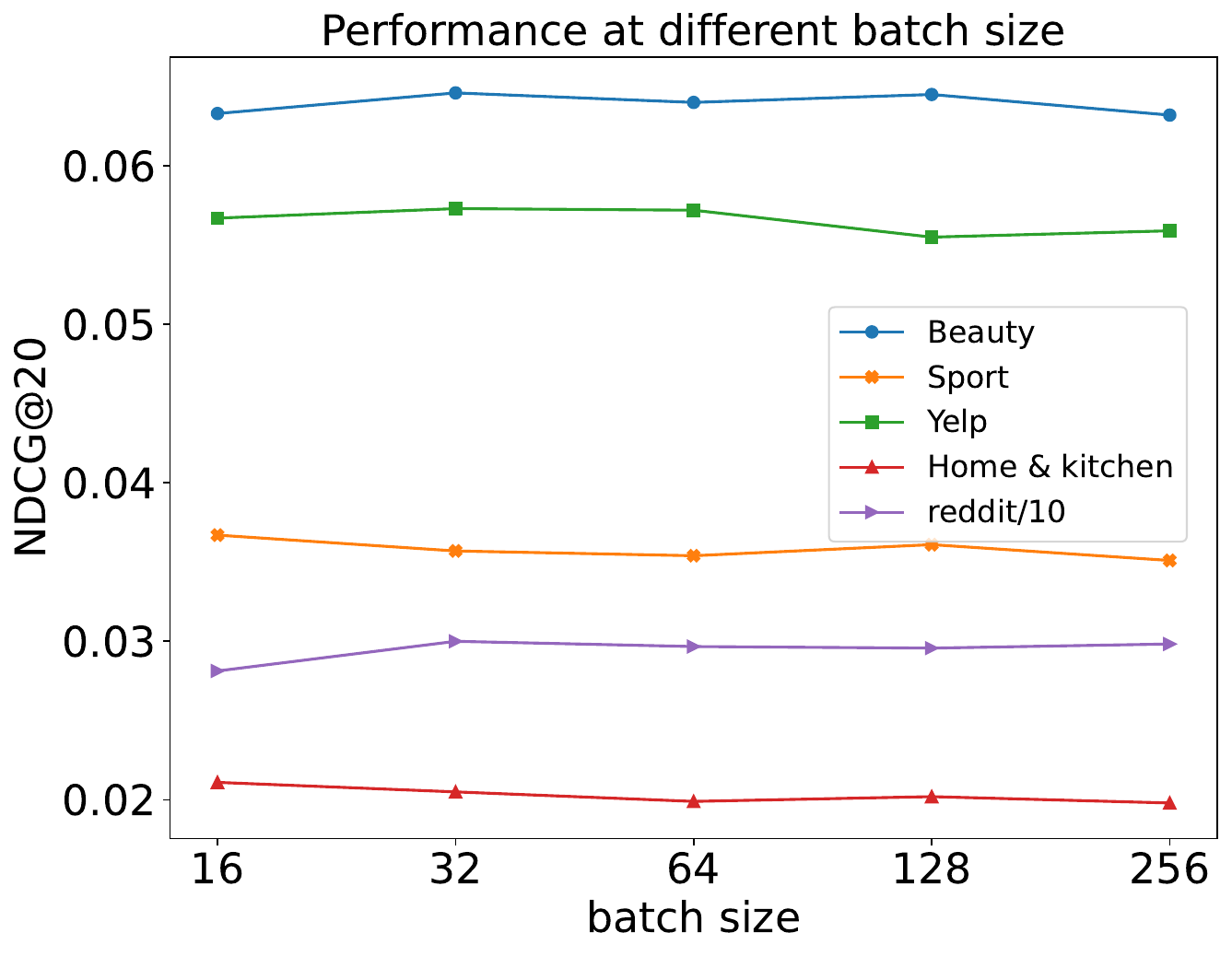}
  \caption{Performance with different batch sizes under NDCG@20. The results from Reddit have been divided by 10 to ensure its line fits to the same figure with the other four datasets.}
  \label{fig:6}
\end{figure}

\begin{figure}[htb]
  \centering
  \includegraphics[width=0.85\linewidth]{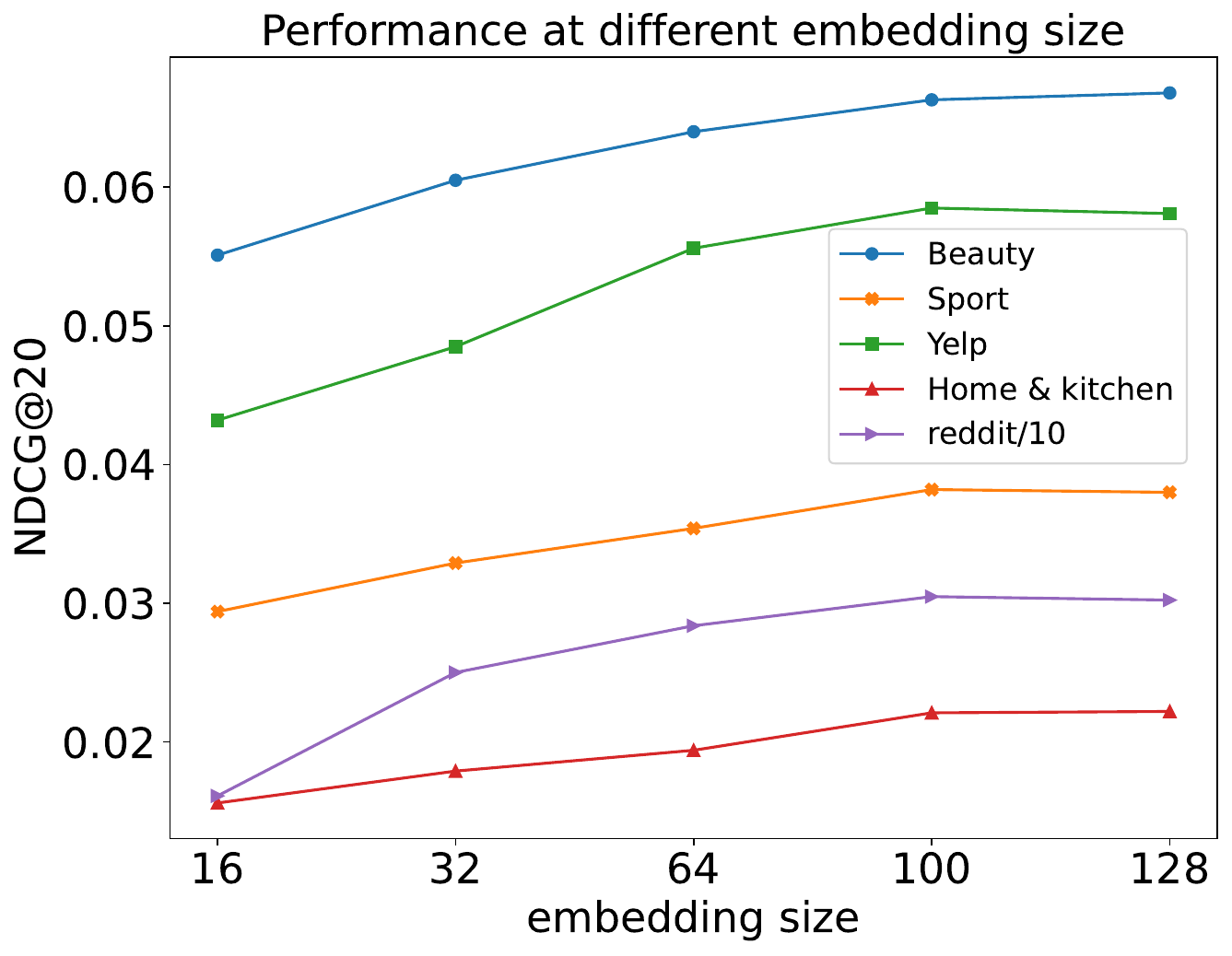}
  \caption{Performance with different embedding sizes under NDCG@20. The results from Reddit have been divided by 10 to ensure its line fits to the same figure with the other four datasets.}
  \label{fig:7}
\end{figure}

\subsubsection{Hyperparameter Analysis on batch size and embedding size}\par

In this section, we explore how a batch size and an embedding size impact the performance of our MVCrec model. We evaluate MVCrec across five datasets using various batch and embedding sizes. NDCG@20 serves as the main metric, similar to the previous section. The batch size ranges from 16 to 256. Results are illustrated in Figure \ref{fig:6}. Optimal batch size varies by each dataset: 32 for Beauty, 16 for Sports, 32 for Yelp, 16 for Home \& Kitchen, and 32 for Reddit. 

The embedding size ranges from 16 to 128. Results are illustrated in Figure \ref{fig:7}. We observe that larger embedding size generally enhances performance across all datasets. Although we do not report HR@20 because of the limited space, we observe that HR@20 has the same trend as NDCG@20 in these experiments. 

\section{Conclusion}
In this paper, we have proposed a novel contrastive learning framework. Our contrastive learning strategy integrated contrastive learning from two views (i.e., item-based sequence and graph-based sequence), enabling our model to learn better sequence representations. To combine the representations extracted from the two views, we employ the concept of multi-view attention fusion method, to generate/learn more expressive sequence representations. Extensive experiments across five benchmark datasets demonstrated the superiority of our model. In this work, we only used the sequence of consumed items without considering the actual time span between them. In the future, we will explore other potential contrastive learning methods based on the temporal sequence to learn even better user and item representations. 
\bibliographystyle{IEEEtran}
\bibliography{IEEEfull}

@inproceedings{wu2022multi,
  title={Multi-view multi-behavior contrastive learning in recommendation},
  author={Wu, Yiqing and Xie, Ruobing and Zhu, Yongchun and Ao, Xiang and Chen, Xin and Zhang, Xu and Zhuang, Fuzhen and Lin, Leyu and He, Qing},
  booktitle={International conference on database systems for advanced applications},
  pages={166--182},
  year={2022},
  organization={Springer}
}

@inproceedings{wang2022multi,
  title={Multi-level contrastive learning framework for sequential recommendation},
  author={Wang, Ziyang and Liu, Huoyu and Wei, Wei and Hu, Yue and Mao, Xian-Ling and He, Shaojian and Fang, Rui and Chen, Dangyang},
  booktitle={Proceedings of the 31st ACM International Conference on Information \& Knowledge Management},
  pages={2098--2107},
  year={2022}
}

@inproceedings{kang2018self,
  title={Self-attentive sequential recommendation},
  author={Kang, Wang-Cheng and McAuley, Julian},
  booktitle={ICDM},
  year={2018},
}

@inproceedings{sun2019bert4rec,
  title={BERT4Rec: Sequential recommendation with bidirectional encoder representations from transformer},
  author={Sun, Fei and Liu, Jun and Wu, Jian and Pei, Changhua and Lin, Xiao and Ou, Wenwu and Jiang, Peng},
  booktitle={CIKM},
  year={2019}
}

@inproceedings{xie2022contrastive,
  title={Contrastive learning for sequential recommendation},
  author={Xie, Xu and Sun, Fei and Liu, Zhaoyang and Wu, Shiwen and Gao, Jinyang and Zhang, Jiandong and Ding, Bolin and Cui, Bin},
  booktitle={ICDE},
  year={2022},
}

@inproceedings{Qiu_2022,
	year = 2022,
 
	author = {Ruihong Qiu and Zi Huang and Hongzhi Yin and Zijian Wang},
  
	title = {Contrastive Learning for Representation Degeneration Problem in Sequential Recommendation},
  
	booktitle = {WSDM}
}

@inproceedings{Qin_2023,
 
	year = 2023,
 
  
	author = {Xiuyuan Qin and Huanhuan Yuan and Pengpeng Zhao and Junhua Fang and Fuzhen Zhuang and Guanfeng Liu and Yanchi Liu and Victor Sheng},
  
	title = {Meta-optimized Contrastive Learning for Sequential Recommendation},
  
	booktitle = {SIGIR}
}

@inproceedings{he2023candidate,
  title={Candidate-aware Graph Contrastive Learning for Recommendation},
  author={He, Wei and Sun, Guohao and Lu, Jinhu and Fang, Xiu Susie},
  booktitle={Proceedings of the 46th International ACM SIGIR Conference on Research and Development in Information Retrieval},
  pages={1670--1679},
  year={2023}
}

@inproceedings{garcin2013personalized,
  title={Personalized news recommendation with context trees},
  author={Garcin, Florent and Dimitrakakis, Christos and Faltings, Boi},
  booktitle={Proceedings of the 7th ACM Conference on Recommender Systems},
  pages={105--112},
  year={2013}
}

@inproceedings{feng2015personalized,
  title={Personalized ranking metric embedding for next new poi recommendation},
  author={Feng, Shanshan and Li, Xutao and Zeng, Yifeng and Cong, Gao and Chee, Yeow Meng},
  booktitle={IJCAI'15 Proceedings of the 24th International Conference on Artificial Intelligence},
  pages={2069--2075},
  year={2015},
  organization={ACM}
}

@inproceedings{tang2018personalized,
  title={Personalized top-n sequential recommendation via convolutional sequence embedding},
  author={Tang, Jiaxi and Wang, Ke},
  booktitle={WSDM},
  year={2018}
}

@inproceedings{Wang_2019,
 
	year = 2019,
 
  
	author = {Xiang Wang and Xiangnan He and Meng Wang and Fuli Feng and Tat-Seng Chua},
  
	title = {Neural Graph Collaborative Filtering},
  
	booktitle = {SIGIR}
}

@inproceedings{Ye_2023,

  
	year = 2023,
  
	author = {Yaowen Ye and Lianghao Xia and Chao Huang},
  
	title = {Graph Masked Autoencoder for Sequential Recommendation},
  
	booktitle = {SIGIR}
}

@inproceedings{chen2020revisiting,
  title={Revisiting graph based collaborative filtering: A linear residual graph convolutional network approach},
  author={Chen, Lei and Wu, Le and Hong, Richang and Zhang, Kun and Wang, Meng},
  booktitle={Proceedings of the AAAI conference on artificial intelligence},
  volume={34},
  number={01},
  pages={27--34},
  year={2020}
}

@inproceedings{Chen_2022,
  
	year = 2022,
  

	author = {Yongjun Chen and Zhiwei Liu and Jia Li and Julian McAuley and Caiming Xiong},
  
	title = {Intent Contrastive Learning for Sequential Recommendation},
  
	booktitle = {WebConf}
}

@article{liu2021contrastive,
  title={Contrastive self-supervised sequential recommendation with robust augmentation},
  author={Liu, Zhiwei and Chen, Yongjun and Li, Jia and Yu, Philip S and McAuley, Julian and Xiong, Caiming},
  journal={arXiv preprint arXiv:2108.06479},
  year={2021}
}

@article{liu2022improving,
  title={Improving contrastive learning with model augmentation},
  author={Liu, Zhiwei and Chen, Yongjun and Li, Jia and Luo, Man and Yu, Philip S and Xiong, Caiming},
  journal={arXiv preprint arXiv:2203.15508},
  year={2022}
}

@article{wang2023sequential,
  title={Sequential recommendation with multiple contrast signals},
  author={Wang, Chenyang and Ma, Weizhi and Chen, Chong and Zhang, Min and Liu, Yiqun and Ma, Shaoping},
  journal={ACM Transactions on Information Systems},
  volume={41},
  number={1},
  pages={1--27},
  year={2023},
  publisher={ACM New York, NY}
}

@InProceedings{Wu_2018_CVPR,
author = {Wu, Zhirong and Xiong, Yuanjun and Yu, Stella X. and Lin, Dahua},
title = {Unsupervised Feature Learning via Non-Parametric Instance Discrimination},
booktitle = {Proceedings of the IEEE Conference on Computer Vision and Pattern Recognition (CVPR)},
month = {June},
year = {2018}
}

@article{oord2018representation,
  title={Representation learning with contrastive predictive coding},
  author={Oord, Aaron van den and Li, Yazhe and Vinyals, Oriol},
  journal={arXiv preprint arXiv:1807.03748},
  year={2018}
}

@inproceedings{he2020momentum,
  title={Momentum contrast for unsupervised visual representation learning},
  author={He, Kaiming and Fan, Haoqi and Wu, Yuxin and Xie, Saining and Girshick, Ross},
  booktitle={Proceedings of the IEEE/CVF conference on computer vision and pattern recognition},
  pages={9729--9738},
  year={2020}
}

@inproceedings{chen2020simple,
  title={A simple framework for contrastive learning of visual representations},
  author={Chen, Ting and Kornblith, Simon and Norouzi, Mohammad and Hinton, Geoffrey},
  booktitle={International conference on machine learning},
  pages={1597--1607},
  year={2020},
  organization={PMLR}
}

@article{grill2020bootstrap,
  title={Bootstrap your own latent-a new approach to self-supervised learning},
  author={Grill, Jean-Bastien and Strub, Florian and Altch{\'e}, Florent and Tallec, Corentin and Richemond, Pierre and Buchatskaya, Elena and Doersch, Carl and Avila Pires, Bernardo and Guo, Zhaohan and Gheshlaghi Azar, Mohammad and others},
  journal={Advances in neural information processing systems},
  volume={33},
  pages={21271--21284},
  year={2020}
}

@inproceedings{yang2023debiased,
  title={Debiased Contrastive Learning for Sequential Recommendation},
  author={Yang, Yuhao and Huang, Chao and Xia, Lianghao and Huang, Chunzhen and Luo, Da and Lin, Kangyi},
  booktitle={Proceedings of the ACM Web Conference 2023},
  pages={1063--1073},
  year={2023}
}

@inproceedings{liu2023joint,
  title={Joint Internal Multi-Interest Exploration and External Domain Alignment for Cross Domain Sequential Recommendation},
  author={Liu, Weiming and Zheng, Xiaolin and Chen, Chaochao and Su, Jiajie and Liao, Xinting and Hu, Mengling and Tan, Yanchao},
  booktitle={WebConf},
  year={2023}
}

@inproceedings{liang2023mmmlp,
  title={MMMLP: Multi-modal Multilayer Perceptron for Sequential Recommendations},
  author={Liang, Jiahao and Zhao, Xiangyu and Li, Muyang and Zhang, Zijian and Wang, Wanyu and Liu, Haochen and Liu, Zitao},
  booktitle={WebConf},
  year={2023}
}

@inproceedings{li2023automlp,
  title={AutoMLP: Automated MLP for Sequential Recommendations},
  author={Li, Muyang and Zhang, Zijian and Zhao, Xiangyu and Wang, Wanyu and Zhao, Minghao and Wu, Runze and Guo, Ruocheng},
  booktitle={WebConf},
  year={2023}
}

@inproceedings{Lin_2023, 
	year = 2023,
	month = {apr},
  
	author = {Guanyu Lin and Chen Gao and Yu Zheng and Jianxin Chang and Yanan Niu and Yang Song and Zhiheng Li and Depeng Jin and Yong Li},
  
	title = {Dual-interest Factorization-heads Attention for Sequential Recommendation},
  
	booktitle = {WebConf}
}

@inproceedings{huang2023modeling,
  title={Modeling Temporal Positive and Negative Excitation for Sequential Recommendation},
  author={Huang, Chengkai and Wang, Shoujin and Wang, Xianzhi and Yao, Lina},
  booktitle={Proceedings of the ACM Web Conference 2023},
  pages={1252--1263},
  year={2023}
}

@inproceedings{fan2022sequential,
  title={Sequential recommendation via stochastic self-attention},
  author={Fan, Ziwei and Liu, Zhiwei and Wang, Yu and Wang, Alice and Nazari, Zahra and Zheng, Lei and Peng, Hao and Yu, Philip S},
  booktitle={Proceedings of the ACM Web Conference 2022},
  pages={2036--2047},
  year={2022}
}

@inproceedings{lin2022sequential,
  title={Sequential Recommendation with Decomposed Item Feature Routing},
  author={Lin, Kun and Wang, Zhenlei and Shen, Shiqi and Wang, Zhipeng and Chen, Bo and Chen, Xu},
  booktitle={Proceedings of the ACM Web Conference 2022},
  pages={2288--2297},
  year={2022}
}

@inproceedings{hou2023learning,
  title={Learning vector-quantized item representation for transferable sequential recommenders},
  author={Hou, Yupeng and He, Zhankui and McAuley, Julian and Zhao, Wayne Xin},
  booktitle={Proceedings of the ACM Web Conference 2023},
  pages={1162--1171},
  year={2023}
}

@inproceedings{wang2020fine,
  title={Fine-grained interest matching for neural news recommendation},
  author={Wang, Heyuan and Wu, Fangzhao and Liu, Zheng and Xie, Xing},
  booktitle={ACL},
  year={2020}
}

@inproceedings{Yang_2022,
	year = 2022,

	author = {Yuhao Yang and Chao Huang and Lianghao Xia and Yuxuan Liang and Yanwei Yu and Chenliang Li},
  
	title = {Multi-Behavior Hypergraph-Enhanced Transformer for Sequential Recommendation},
  
	booktitle = {KDD}
}

@inproceedings{li2023exploiting,
  title={Exploiting Explicit and Implicit Item relationships for Session-based Recommendation},
  author={Li, Zihao and Wang, Xianzhi and Yang, Chao and Yao, Lina and McAuley, Julian and Xu, Guandong},
  booktitle={Proceedings of the Sixteenth ACM International Conference on Web Search and Data Mining},
  pages={553--561},
  year={2023}
}

@inproceedings{he2022query,
  title={Query-Aware Sequential Recommendation},
  author={He, Zhankui and Zhao, Handong and Wang, Zhaowen and Lin, Zhe and Kale, Ajinkya and Mcauley, Julian},
  booktitle={Proceedings of the 31st ACM International Conference on Information \& Knowledge Management},
  pages={4019--4023},
  year={2022}
}

@inproceedings{wei2023multi,
  title={Multi-Modal Self-Supervised Learning for Recommendation},
  author={Wei, Wei and Huang, Chao and Xia, Lianghao and Zhang, Chuxu},
  booktitle={Proceedings of the ACM Web Conference 2023},
  pages={790--800},
  year={2023}
}

@article{zhang2023multimodal,
  title={Multimodal Pre-training Framework for Sequential Recommendation via Contrastive Learning},
  author={Zhang, Lingzi and Zhou, Xin and Shen, Zhiqi},
  journal={arXiv preprint arXiv:2303.11879},
  year={2023}
}

@inproceedings{10.1145/3543507.3583529,
author = {Fan, Ziwei and Liu, Zhiwei and Peng, Hao and Yu, Philip S},
title = {Mutual Wasserstein Discrepancy Minimization for Sequential Recommendation},
year = {2023},
booktitle = {WebConf},

}

@inproceedings{10.1145/3397271.3401063,
author = {He, Xiangnan and Deng, Kuan and Wang, Xiang and Li, Yan and Zhang, YongDong and Wang, Meng},
title = {LightGCN: Simplifying and Powering Graph Convolution Network for Recommendation},
year = {2020},
booktitle = {SIGIR},
}

@inproceedings{DBLP:journals/corr/HidasiKBT15,
  author       = {Bal{\'{a}}zs Hidasi and
                  Alexandros Karatzoglou and
                  Linas Baltrunas and
                  Domonkos Tikk},
  title        = {Session-based Recommendations with Recurrent Neural Networks},
  booktitle    = {ICLR},
  year         = {2016},
}

@inproceedings{10.1145/3580305.3599519,
author = {Li, Jiacheng and Wang, Ming and Li, Jin and Fu, Jinmiao and Shen, Xin and Shang, Jingbo and McAuley, Julian},
title = {Text Is All You Need: Learning Language Representations for Sequential Recommendation},
year = {2023},
booktitle = {KDD},

}

@inproceedings{10.1145/3543507.3583479,
author = {Lin, Yujie and Wang, Chenyang and Chen, Zhumin and Ren, Zhaochun and Xin, Xin and Yan, Qiang and de Rijke, Maarten and Cheng, Xiuzhen and Ren, Pengjie},
title = {A Self-Correcting Sequential Recommender},
year = {2023},
booktitle = {WebConf},

}

@inproceedings{10.1145/3459637.3482291,
author = {Mao, Kelong and Zhu, Jieming and Xiao, Xi and Lu, Biao and Wang, Zhaowei and He, Xiuqiang},
title = {UltraGCN: Ultra Simplification of Graph Convolutional Networks for Recommendation},
year = {2021},





booktitle = {CIKM},

}

@inproceedings{10.5555/1795114.1795167,
author = {Rendle, Steffen and Freudenthaler, Christoph and Gantner, Zeno and Schmidt-Thieme, Lars},
title = {BPR: Bayesian Personalized Ranking from Implicit Feedback},
year = {2009},
isbn = {9780974903958},
publisher = {AUAI Press},
address = {Arlington, Virginia, USA},
booktitle = {Proceedings of the Twenty-Fifth Conference on Uncertainty in Artificial Intelligence},
pages = {452–461},
numpages = {10},
location = {Montreal, Quebec, Canada},
series = {UAI '09}
}

@inproceedings{10.1145/3539618.3591691,
author = {Yang, Yonghui and Wu, Zhengwei and Wu, Le and Zhang, Kun and Hong, Richang and Zhang, Zhiqiang and Zhou, Jun and Wang, Meng},
title = {Generative-Contrastive Graph Learning for Recommendation},
year = {2023},





booktitle = {SIGIR},
}

@article{Wu_2019,
   title={Session-Based Recommendation with Graph Neural Networks},
   volume={33},
   number={01},
   journal={Proceedings of the AAAI Conference on Artificial Intelligence},
   publisher={Association for the Advancement of Artificial Intelligence (AAAI)},
   author={Wu, Shu and Tang, Yuyuan and Zhu, Yanqiao and Wang, Liang and Xie, Xing and Tan, Tieniu},
   year={2019},
   month=jul, pages={346–353} }

@misc{du2023ensemble,
      title={Ensemble Modeling with Contrastive Knowledge Distillation for Sequential Recommendation}, 
      author={Hanwen Du and Huanhuan Yuan and Pengpeng Zhao and Fuzhen Zhuang and Guanfeng Liu and Lei Zhao and Victor S. Sheng},
      year={2023},
      eprint={2304.14668},
      archivePrefix={arXiv},
      primaryClass={cs.IR}
}

@inproceedings{10.1145/3539618.3591717,
author = {Liu, Langming and Cai, Liu and Zhang, Chi and Zhao, Xiangyu and Gao, Jingtong and Wang, Wanyu and Lv, Yifu and Fan, Wenqi and Wang, Yiqi and He, Ming and Liu, Zitao and Li, Qing},
title = {LinRec: Linear Attention Mechanism for Long-Term Sequential Recommender Systems},
year = {2023},
booktitle = {SIGIR},
}

@inproceedings{Kim_2023, series={SIGIR ’23},
   title={MELT: Mutual Enhancement of Long-Tailed User and Item for Sequential Recommendation},
   booktitle={SIGIR},
   author={Kim, Kibum and Hyun, Dongmin and Yun, Sukwon and Park, Chanyoung},
   year={2023},
}

@inproceedings{li2024category,
  title={Category-based and popularity-guided video game recommendation: a balance-oriented framework},
  author={Li, Xiping and Ma, Jianghong and Liu, Kangzhe and Feng, Shanshan and Zhang, Haijun and Wang, Yutong},
  booktitle={Proceedings of the ACM Web Conference 2024},
  pages={3734--3744},
  year={2024}
}

@article{li2026cpgrec+,
  title={CPGRec+: A Balance-oriented Framework for Personalized Video Game Recommendations},
  author={Li, Xiping and Yang, Aier and Ma, Jianghong and Liu, Kangzhe and Feng, Shanshan and Zhang, Haijun and Zhao, Yi},
  journal={ACM Transactions on Information Systems},
  volume={44},
  number={3},
  pages={1--44},
  year={2026},
  publisher={ACM New York, NY}
}

\end{document}